\documentclass{jpsj-suppl}
\usepackage{txfonts} 

\title{Field-induced Phenomena in Ferromagnetic Superconductors UCoGe and URhGe}

\author{
Dai~\textsc{Aoki}$^1$, 
Mathieu~\textsc{Taupin}$^1$,
Carley~\textsc{Paulsen}$^2$,
Fr\'{e}d\'{e}ric~\textsc{Hardy}$^3$,
Valentin~\textsc{Taufour}$^1$,
Hisashi~\textsc{Kotegawa}$^{4,1}$,
Elena~\textsc{Hassinger}$^1$,
Liam~\textsc{Malone}$^{5,8,1}$,
Tatsuma~D.~\textsc{Matsuda}$^{6,1}$,
Atsushi~\textsc{Miyake}$^{7,1}$, 
Ilya~\textsc{Sheikin}$^5$,
William~\textsc{Knafo}$^8$,
Georg~\textsc{Knebel}$^1$,
Ludovic~\textsc{Howald}$^1$,
Jean-Pascal~\textsc{Brison}$^1$ and
Jacques~\textsc{Flouquet}$^1$
}

\inst{%
$^1$INAC/SPSMS, CEA, 38054 Grenoble, France\\
$^2$Institut N\'{e}el, CNRS \& Universit\'{e} Joseph Fourier, BP 166, 38042 Grenoble, France\\
$^3$KIT, Institut f\"{u}r Festk\"{o}rperphysik, 76021 Karlsruhe, Germany\\
$^4$Department of Physics, Kobe University, Kobe 657-8501, Japan\\
$^5$LNCMI-G, CNRS, 38042 Grenoble, France\\
$^6$ASRC, JAEA, Tokai, Ibaraki 319-1195, Japan\\
$^7$KYOKUGEN, Osaka University, Toyonaka, Osaka 560-8531, Japan\\
$^8$LNCMI-T, CNRS, 31400 Toulouse, France
}

\email{dai.aoki@cea.fr}

\recdate{}

\abst{
We review our recent studies on ferromagnetic superconductors, UGe$_2$, URhGe and UCoGe,
where the spin-triplet state with the so-called equal spin pairing is realized.
We focus on experimental results of URhGe and UCoGe in which the superconductivity occurs already at ambient pressure.
The huge upper critical field $H_{\rm c2}$ on UCoGe for the field along the hard magnetization axis ($b$-axis) is confirmed by the AC susceptibility measurements by the fine tuning of field angle.
Contrary to the huge $H_{\rm c2}$ along the hard-magnetization axis, $H_{\rm c2}$ along the easy-magnetization axis ($c$-axis) is relatively small in value.
However, the initial slope of $H_{\rm c2}$, namely $dH_{\rm c2}/dT$ ($H\to 0$) both in UCoGe and in URhGe indicates the large value, which can be explained by the magnetic domain effect detected in the magnetization measurements.
The specific heat measurements using a high quality single crystal of UCoGe demonstrate 
the bulk superconductivity, which is extended under magnetic field for the field along $c$-axis.
}

\kword{superconductivity, ferromagnetism, uranium compound, superconducting upper critical field, AC susceptibility, specific heat, resistivity, magnetization}

\begin{document}
\maketitle

Ferromagnetism and superconductivity had been thought to be competitive phenomena, since the large internal field due to the ferromagnetism easily destroys the superconducting Cooper pairs. 
In 1980s', some exceptional cases were found for example in ErRh$_4$B$_4$ and HoMo$_6$Se$_8$~\cite{Fis90_FM}.
In these materials, the superconductivity is observed in the intermediate temperature range.
Further decreasing temperature, the ferromagnetism appears and the superconducting phase collapses at low temperature, meaning that the Curie temperature ($T_{\rm Curie}$) is lower than the superconducting transition temperature ($T_{\rm sc}$), and there is no microscopic coexistence of superconductivity and ferromagnetism with large ordered moment.
The microscopic coexistence was theoretically proposed in the weak ferromagnet ZrZn$_2$~\cite{Fay80},
where the spin-triplet state is formed near the ferromagnetic quantum critical point.
There are, however, no experimental evidences.

The first microscopic coexistence of ferromagnetism and superconductivity was found in UGe$_2$ under pressure~\cite{Sax00}, then in URhGe at ambient pressure~\cite{Aok01}, and more recently in UCoGe at ambient pressure~\cite{Huy07}.
All the known ferromagnetic superconductors are uranium compounds, and $T_{\rm sc}$ is lower than $T_{\rm Curie}$~\cite{Aok11_CR,Aok11_JPSJ_review}.
The ordered moments of U atom are substantially reduced, compared to the free ion value. 
(UGe$_2$: $1.5\,\mu_{\rm B}$,
 URhGe:   $0.4\,\mu_{\rm B}$,
 UCoGe:   $0.05\,\mu_{\rm B}$)
The Sommerfeld coefficients ($\gamma$-value) are moderately enhanced
(UGe$_2$: $34\,{\rm mJ\,K^{-2}mol^{-1}}$,
 URhGe:   $160\,{\rm mJ\,K^{-2}mol^{-1}}$,
 UCoGe:   $55\,{\rm mJ\,K^{-2}mol^{-1}}$),
indicating the proximity of the ferromagnetic quantum critical point.
A striking point is the large upper critical field $H_{\rm c2}$ exceeding the Pauli paramagnetic limit.
When the field is applied along the hard-magnetization axis ($b$-axis) in URhGe, 
the field-reentrant superconducting phase appears~\cite{Lev05}.
In UCoGe, the unusual S-shaped $H_{\rm c2}$ curve is observed~\cite{Aok09_UCoGe}.
These phenomena are closely related to the ferromagnetic instabilities,
where $T_{\rm Curie}$ is reduced under magnetic fields and collapses 
at the enhanced superconducting phase.
Another interesting point is that $T_{\rm sc}$ of UCoGe shows the maximum at the critical pressure $P_{\rm c}$ where ferromagnetism is suppressed,
and superconductivity survives even in the paramagnetic state~\cite{Has08_UCoGe}.
This is contradictory to a previous theoretical prediction~\cite{Fay80}.

Here we present our recent experimental results on UCoGe and URhGe for further investigation on the superconducting state and the ferromagnetic instabilities.
High quality single crystals of UCoGe and URhGe were grown using the Czochralski method in a tetra-arc furnace. 
Starting materials with appropriate ratio were melted under the high purity Ar atmosphere gas.
The ingot was then pulled using a W-tip or a seed crystal with a slow pulling rate $10$--$15\,{\rm mm/hr}$.
Obtained single crystals were cut in a spark cutter and were annealed under the ultra high vacuum.
The quality of single crystals characterized by resistivity measurements strongly depends on the position of the single crystal ingot. 
Therefore strenuous efforts were devoted for the characterization.
For example, more than 150 single crystal samples of UCoGe were checked by the resistivity measurements using a homemade adiabatic demagnetization refrigerator (ADR) combined with PPMS, 
which allows us to cool the samples down to $100\,{\rm mK}$ within two hours from room temperature.
The highest residual resistivity ratio ($RRR$) of UCoGe is 200.
The temperature dependence of AC susceptibility of UCoGe was measured at low temperature down to $0.1\,{\rm K}$ and at high field up to $16\,{\rm T}$ for the field along $b$-axis using a conventional dilution refrigerator with a sample rotation system.
The specific heat of UCoGe was measured by the relaxation method using a homemade addenda at low temperature down to $0.1\,{\rm K}$ and at fields up to $0.5\,{\rm T}$ for the field along $c$-axis.
The very low field resistivity measurements for the field along $c$-axis were performed in UCoGe. 
Special attention was paid in order to eliminate the remnant field using a Hall sensor.
The low field magnetization measurements were done for the field along $c$-axis by using a homemade SQUID magnetometer for UCoGe and a commercial SQUID magnetometer for URhGe.

\begin{figure}[tbh]
\begin{center}
\includegraphics[width=0.8 \hsize,clip]{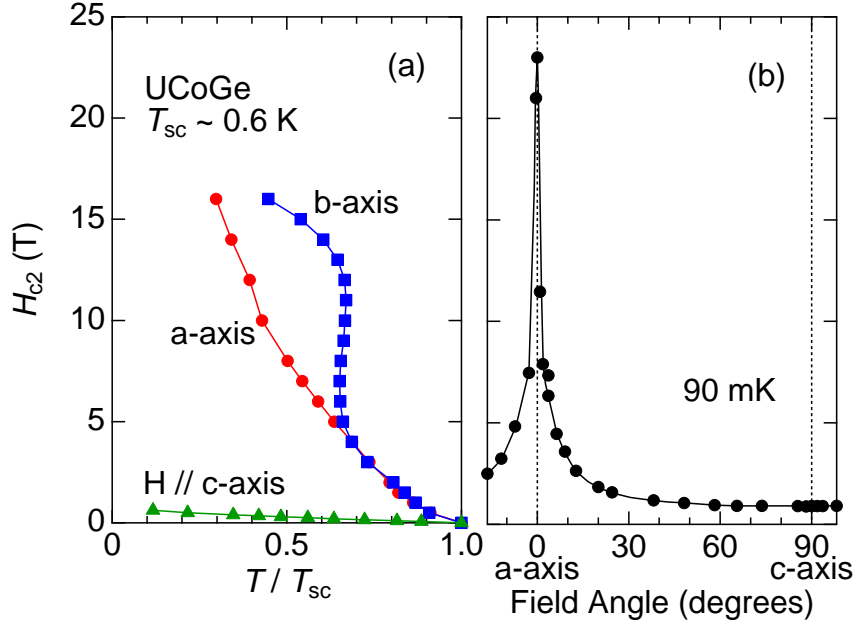}
\end{center}
\caption{(a)Temperature dependence of the upper critical field $H_{\rm c2}$ for $H\parallel a$, $b$ and $c$-axes determined by the resistivity measurements in UCoGe. $T_{\rm sc}$ at zero field is approximately $0.6\,{\rm K}$, which slightly depends on the sample. (b)Angular dependence of $H_{\rm c2}$ at $90\,{\rm mK}$~\cite{Aok09_UCoGe}.}
\label{fig:UCoGe_Hc2}
\end{figure}
Figure~\ref{fig:UCoGe_Hc2}(a) shows the $H_{\rm c2}$ curves of UCoGe for $H\parallel a$, $b$ and $c$-axes
determined by the resistivity measurements.
The value of $H_{\rm c2}$ for $H \parallel c$-axis is close to the Pauli limit based on the weak coupling BCS model.
When the field is applied along the hard magnetization axis ($a$ and $b$-axis),
the huge $H_{\rm c2}$ are observed.
For $H\parallel a$-axis, $H_{\rm c2}$ shows the upward curvature and may reaches $\sim 30\,{\rm T}$ at $0\,{\rm K}$.
A very anisotropic $H_{\rm c2}$ between $a$ and $c$-axis is also displayed in Fig.~\ref{fig:UCoGe_Hc2}(b).
When the field is slightly tilted to the easy magnetization axis, $H_{\rm c2}$ is strongly suppressed.
$H_{\rm c2}$ for $H\parallel b$-axis shows the unusual S-shaped curve.
Since $T_{\rm Curie}$ is suppressed from $2.5\,{\rm K}$ at zero field to $1\,{\rm K}$ at $14\,{\rm T}$ for $H\parallel b$-axis, one can consider that the ferromagnetic instabilities induced by the magnetic field reinforce the superconducting state.
In fact, the resistivity $A$ coefficient and the $\gamma$-value are enhanced around $14\,{\rm T}$~\cite{Aok09_UCoGe,Aok11_ICHE}.
The situation seems to be similar to the case of URhGe, where the spin-reorientation, re-entrant superconductivity, enhancement of $\gamma$-value are observed around $H_{\rm m}\sim 12\,{\rm T}$~\cite{Lev05,Miy08,Har11,Aok11_ICHE}.
However, the paradox of UCoGe is that no anomaly is detected around $14\,{\rm T}$ in the magnetization measurements up to now.
A new mechanism associated with the reconstruction of Fermi surface is also plausible in UCoGe.
In fact, non-linear field-response of Shubnikov de Haas frequency is observed above $20\,{\rm T}$~\cite{Aok11_UCoGe},
implying the Fermi surface instabilities.
Furthermore, since UCoGe is low carrier system, the Fermi surface can be easily affected by the magnetic field as in URu$_2$Si$_2$~\cite{Shi09,Has10_URu2Si2}.

\begin{figure}[tbh]
\begin{center}
\includegraphics[width= \hsize,clip]{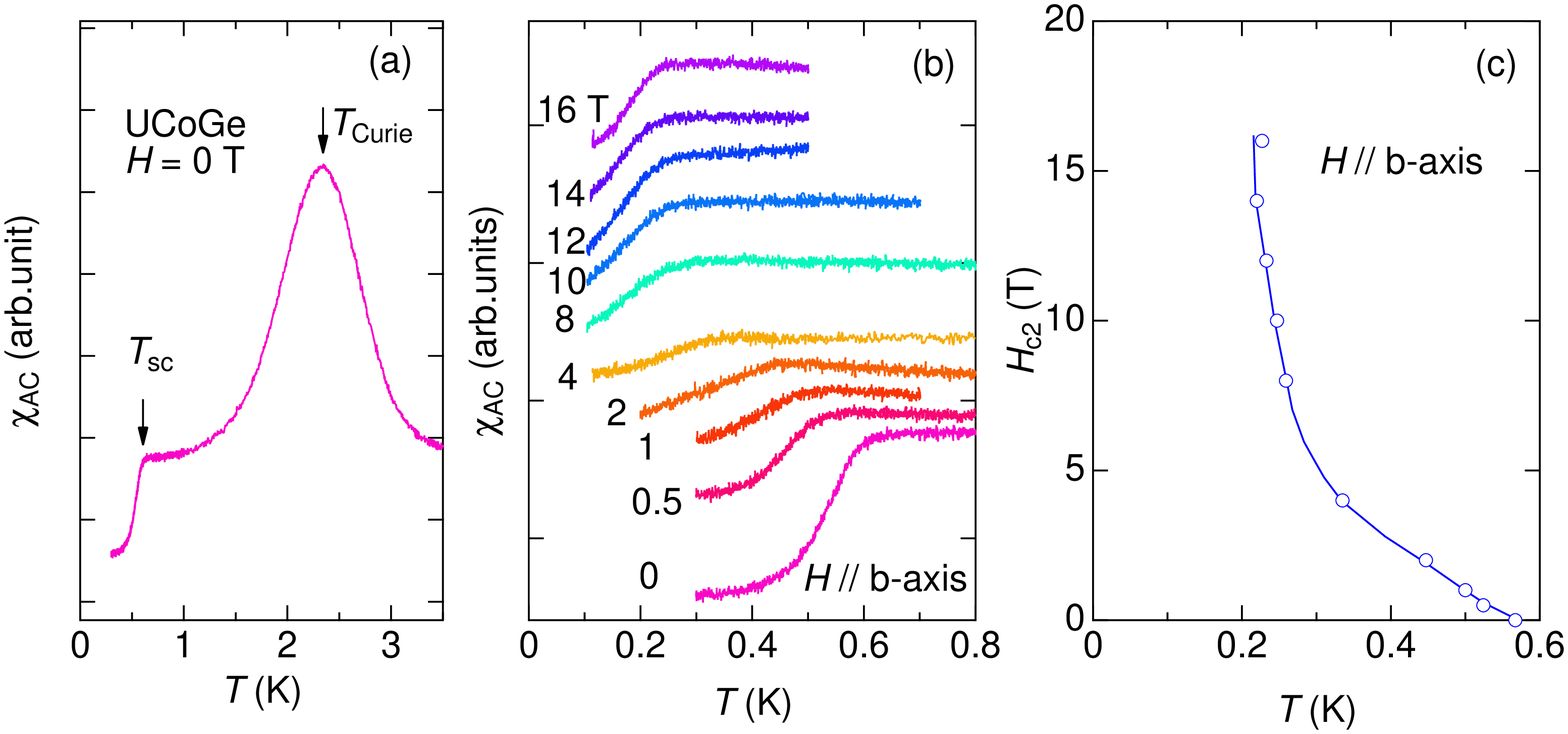}
\end{center}
\caption{Temperature dependence of the AC susceptibility at zero field (a) and under magnetic field at low temperature for $H\parallel b$-axis (b) in UCoGe. (c) Temperature dependence of $H_{\rm c2}$ for $H \parallel b$-axis obtained by AC susceptibility measurements.}
\label{fig:UCoGe_AC_chi}
\end{figure}
Figure~\ref{fig:UCoGe_AC_chi} shows the temperature dependence of the AC susceptibilities in UCoGe.
Two anomalies associated with $T_{\rm Curie}$ and $T_{\rm sc}$ are observed at zero field.
This is consistent with the previous report~\cite{Deg10}.
Applying the magnetic field along $b$-axis, the anomaly due to the superconductivity is 
shifted to lower temperatures, but it sustains even at the highest field $16\,{\rm T}$,
indicating the S-shaped $H_{\rm c2}$ curve, as shown in Fig.~\ref{fig:UCoGe_AC_chi}(c),
where the superconducting transition was defined as the onset of the drop of AC susceptibility.
The results are in good agreement with those of resistivity measurements, 
implying the bulk superconductivity for $H\parallel b$-axis,
although another experimental probe such as specific heat or thermal conductivity is required for the definite conclusion.

\begin{figure}[tbh]
\begin{center}
\includegraphics[width= \hsize,clip]{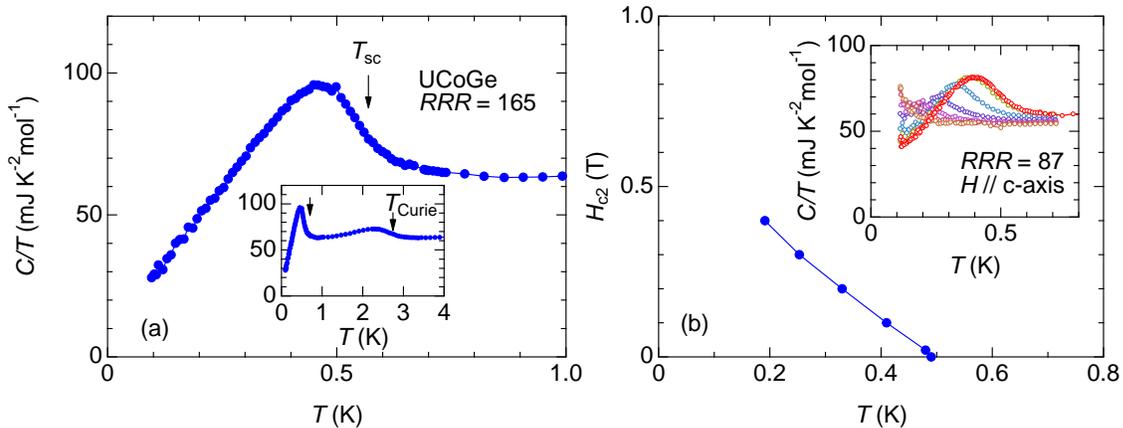}
\end{center}
\caption{(a) Temperature dependence of the specific heat in the form of $C/T$ vs $T$ in the high quality single crystal of UCoGe with $RRR=165$. The inset shows the results at high temperatures. The nuclear specific heat is subtracted. (b) Temperature dependence of $H_{\rm c2}$ for $H\parallel c$-axis obtained by the specific heat. The inset shows the temperature dependence of the specific heat at different fields 0, 0.02, 0.1, 0.2, 0.3, 0.4 and $0.5\,{\rm T}$.}
\label{fig:UCoGe_Cp}
\end{figure}
Figure~\ref{fig:UCoGe_Cp}(a) shows the temperature dependence of the specific heat using a high quality single crystal of UCoGe with $RRR=165$. 
Here the hyperfine contribution due to the nuclear specific heat appeared as an upturn below $0.15\,{\rm K}$
is subtracted, assuming $C_{\rm nuc}\propto 1/T^2$.
Thanks to the very high quality single crystal, the residual $\gamma$-value is relatively small ($\gamma_0\sim 10\,{\rm mJ\,K^{-2}mol^{-1}}$).
This corresponds to $15\,{\%}$ of the specific heat in the normal state ($\gamma_0/\gamma_{\rm N}=0.15$).
Two other ferromagnetic superconductors, URhGe and UGe$_2$ also have residual $\gamma$-values;
$\gamma_0/\gamma_{\rm N}=0.5$ for URhGe with $RRR=120$~\cite{Aok01}, 
$\gamma_0/\gamma_{\rm N}=0.7$ for UGe$_2$ with $RRR=600$~\cite{Tat01} under pressure.
It is worthwhile to note that the residual value seems to be related to the ordered moments
(UGe$_2$: $1\,\mu_{\rm B}$, URhGe: $0.4\,\mu_{\rm B}$ and UCoGe: $0.05\,\mu_{\rm B}$).
The large ordered moment induces the large residual $\gamma$-value.

A reason for the large residual $\gamma$-values could be self-induced vortex states 
caused by the sublattice moment of ferromagnetism.
That is, the material is always in the superconducting mixed state without $H_{\rm c1}$,
which is also supported by recent NMR/NQR experiments and magnetization measurements in UCoGe~\cite{Oht08,Deg10,Pau11_UCoGe}.
Another possibility is that the minority-spin Fermi surface is not gapped and
the majority-spin Fermi surface is only responsible for the superconductivity
within the framework of spin-triplet state with equal-spin pairing.

Figure~\ref{fig:UCoGe_Cp}(b) shows the temperature dependence of $H_{\rm c2}$ for $H\parallel c$-axis,
determined by the specific heat results as a bulk superconductivity.
Because of the large hyperfine contribution at low temperatures under magnetic field as shown in the inset of Fig.~\ref{fig:UCoGe_Cp}(b),
it is difficult to determine $T_{\rm sc}$ under magnetic fields.
Nevertheless, from the entropy balance, $T_{\rm sc}$ was determined up to $0.4\,{\rm T}$, as shown in the main panel of Fig.~\ref{fig:UCoGe_Cp}(b).
The $H_{\rm c2}$ curve exhibits a slight upturn behavior and it may reaches $0.6$--$0.7\,{\rm T}$ at $0\,{\rm K}$.
These results are in good agreement with those determined by the resistivity measurements.

\begin{figure}[tbh]
\begin{center}
\includegraphics[width= \hsize,clip]{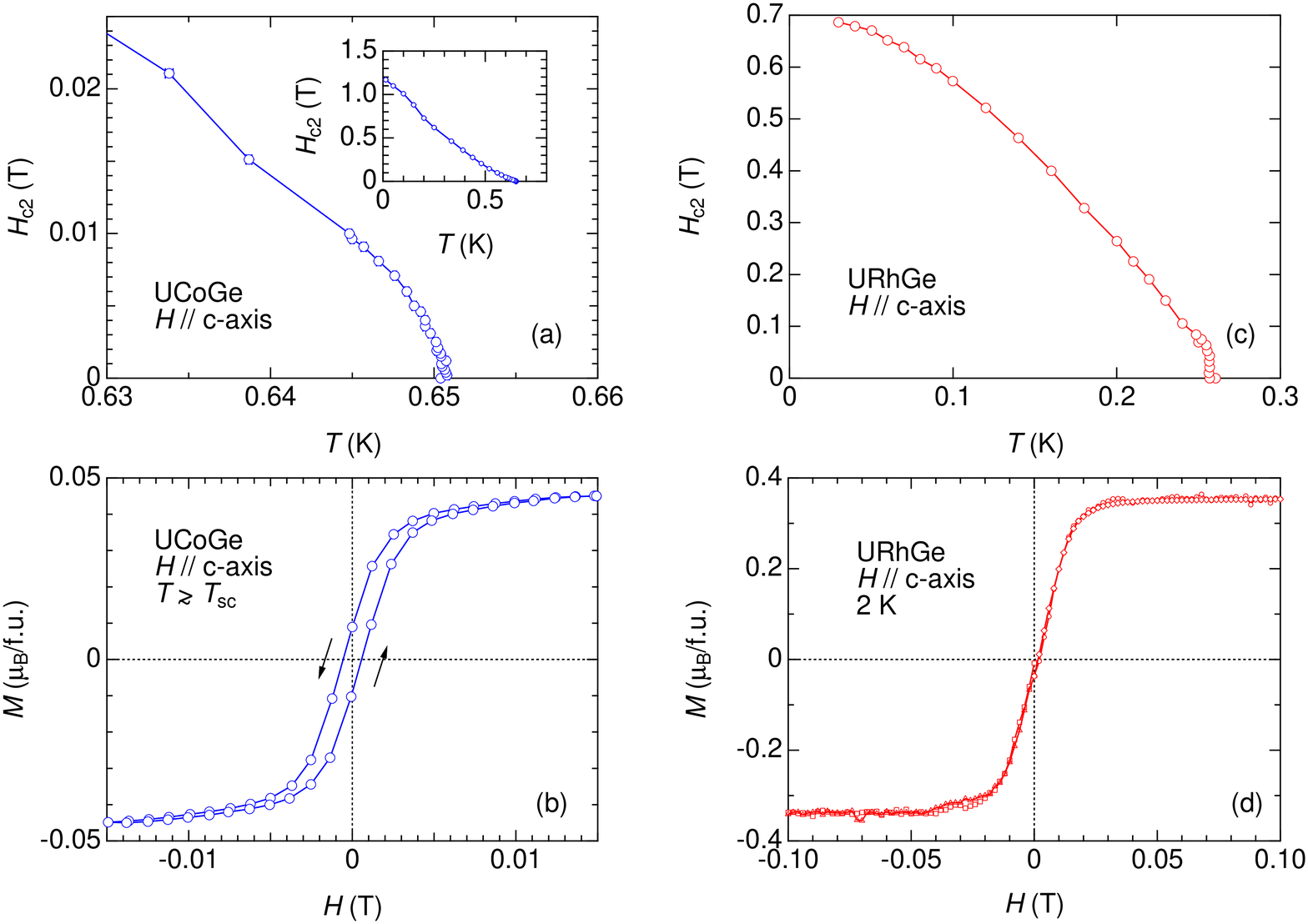}
\end{center}
\caption{(a) Temperature dependence of $H_{\rm c2}$ at low fields determined by the resistivity measurements for $H\parallel c$-axis in UCoGe. The inset shows the full curve of $H_{\rm c2}$. 
(b) Low field magnetization curve just above $T_{\rm sc}$ for $H\parallel c$-axis in UCoGe.
(c) Temperature dependence of $H_{\rm c2}$ for $H\parallel c$-axis in URhGe cited from Ref.~\protect\citen{Har05}. (d) Low field magnetization curve for $H\parallel c$-axis at $2\,{\rm K}$ in URhGe.}
\label{fig:UCoGe_URhGe_Hc2_mag}
\end{figure}
Figures~\ref{fig:UCoGe_URhGe_Hc2_mag}(a) and \ref{fig:UCoGe_URhGe_Hc2_mag}(c) shows the temperature dependence of $H_{\rm c2}$ for $H\parallel c$-axis.
The data of Fig.~\ref{fig:UCoGe_URhGe_Hc2_mag}(c) were cited from Ref.~\citen{Har05}.
Both in UCoGe and in URhGe, unusual initial slopes of $H_{\rm c2}$, which are almost vertical, 
are observed for the field along easy-magnetization axes.
This might be explained by the magnetic domain effect.
As shown in Figs.~\ref{fig:UCoGe_URhGe_Hc2_mag}(b) and \ref{fig:UCoGe_URhGe_Hc2_mag}(d), the magnetic domain is aligned at around $0.005\,{\rm T}$ for UCoGe
and $0.03\,{\rm T}$ for URhGe.
Since the ordered moment of UCoGe is one order magnitude smaller than that of URhGe,
the difference between UCoGe and URhGe with respect to the critical field for the magnetic mono-domain
is consistent.
When the magnetic mono-domain is achieved, $H_{\rm c2}$ decreases with field.
In other words, the initial slope of $H_{\rm c2}$ reveals almost vertical
until the magnetic mono-domain is achieved, because the effective field for superconductivity 
would be almost invariant up to the critical field of magnetic mono-domain.

In summary, we have performed the resistivity, AC-susceptibility, specific heat and magnetization measurements 
in high quality single crystals of UCoGe and URhGe.
The very large $H_{\rm c2}$ was detected by the AC-susceptibility for $H\parallel b$-axis in UCoGe, 
which is consistent with the results of resistivity measurements.
Specific heat in the high quality single crystal of UCoGe ($RRR=165$) reveals the small residual $\gamma$-value in the superconducting state due to the self-induced vortex states which might be related to the magnitude of the ordered moment.
Unusual initial slopes of $H_{\rm c2}$ for the field along the magnetization easy-axis could be explained
by the magnetic domain effect,
indicating an evidence for the strong correlation between ferromagnetism and superconductivity.


This work was supported by ERC starting grant (NewHeavyFermion), French ANR project (CORMAT, SINUS, DELICE).



\end{document}